\documentclass[showpacs,twocolumn,pre,floatfix]{revtex4}
\usepackage[]{graphicx}
\usepackage[]{color}
\usepackage{amsmath}
\usepackage{amssymb}
\usepackage{ulem}

\newcommand{\rb}{{\bf r}}

\newcommand{\Rb}{{\bf R}}
\newcommand{\Fb}{{\bf F}}
\newcommand{\Jb}{{\bf J}}
\newcommand{\Vb}{{\bf V}}

\newcommand{\vct}[1]{\mathbf{#1}}

\begin{document}
\title{Theory of rheology in confinement}

\date{\today}

\author{Artem A. Aerov}
\affiliation{4th Institute for Theoretical Physics, Universit\"at Stuttgart, Germany and Max Planck Institute for Intelligent Systems, 70569 Stuttgart, Germany}
\email{aerov@is.mpg.de}
\author{Matthias Kr\"uger}
\affiliation{4th Institute for Theoretical Physics, Universit\"at Stuttgart, Germany and Max Planck Institute for Intelligent Systems, 70569 Stuttgart, Germany}
\email{aerov@is.mpg.de}

\begin{abstract}
Viscosity of fluids is generally understood in terms of kinetic mechanisms, i.e. particle collisions, or thermodynamic ones as imposed through structural distortions upon
e.g.~applying shear. Often the latter are more relevant, which allows a simpler theoretical description, and e.g. (damped) Brownian particles can be considered good fluid model systems.
We formulate a general theoretical approach for 
rheology in confinement, based on microscopic equations of motion and classical density functional theory. Specifically, we discuss the viscosity for the case of two parallel walls in relative motion as a function of wall-to-wall distance, analyzing its relation to the slip length found for a single wall. The previously observed [{\it J.~Chem.~Phys. }\textbf {140}, {094701} ({2014})] deficiency of inhomogeneous (unphysical) stresses under naive application of shear in confinement is healed when including hydrodynamic interactions.
\end{abstract}


\maketitle

\section{Introduction}
Viscosity of fluids is important for technology and biology. It has been investigated
for many years \cite{Maxwell_viscosity, Viscosity_of_Liquids_0,  ElliottPRL2014ViscFermGasMeasurement0}, e.g. using linear response theory \cite{Green, Kubo}.     

A lot is known about bulk rheology. The response of dilute gases \cite{Andrade} can be analyzed by kinetic theory \cite{Kreuzer}.  
For (Brownian) suspensions insight has been gained e.g. for {dilute \cite{brady_morris} or glassy 
\cite{Fuchs02, Sollich97_0, Falk98, Barrat2013FaradDiscModelChannel} systems, here also nonlinear effects are accessible by theory and by experiment \cite{joe_review}. 
Improved experimental precision on small scales \cite{IgnatovichPRL2006DetectingNanoPart, Isa09_0, Cheng11_0, Chevalier_0}  has boosted also the research of confined systems
\cite{Petravic2_2006_JCP_viscosity_lin_resp_between_planes, Klapp2008PRL_SimCollFilm, Zhang2004SimPoisFl_0},
which is important for e.g. microfluidic devices \cite{Psaltis2006_Nature_microfl_dev_0, Sajeesh2014_rev_Part_sep_in_Mfl_dev}, 
MEMS \cite{HoTaiAnnRevFlMech1998_MEMS_flu_flows, CaoSunIntJMolSci2009MEMS_fluid_review_0} 
or blood flow in capillaries \cite{Li_blood_fl_sim_0, Zhou2005_capillary_blood_susp}.  

Theoreticians have put much effort in describing many body systems \cite{HansenMcDonald}, where successful (approximate) approaches, based on first principles, 
include mode coupling theory \cite{Fuchs02, pnas_0, Miyazaki02} or density functional theory (DFT) \cite{bob_advances, Marconi99, archer}. Using such methods, bulk rheology of dense systems
\cite{pnas_0, Miyazaki02, Fuchs02} or the evolution of density profiles under time varying potentials \cite{Marconi99, archer, RexLoewenPRL2008} have been studied. 
There is also recent progress towards dense driven systems in inhomogeneous situations \cite{Dhont14}. 

We present a  theory of rheology in confinement based on first principles. The exact equations need an 
approximative closure for the two-particle density, and reproduce known results for the limit of inessential confinement. 
Explicitly, we study the case of suspensions,
starting from the Smoluchowski equation of motion \cite{Risken} with hydrodynamic interactions, 
and consider how the effective viscosity between two parallel walls depends on the distance between them. 
We study this scenario by two approaches, first taking into account hydrodynamic interactions, and second using a simplified model, where hydrodynamic interactions are neglected \cite{AerovKrugerJCP2014}. The latter yields a simple relation between the effective viscosity and the previously obtained slip length, 
and reproduces many features observed in simulations of molecular fluids. In contrast to previous approaches   
\cite{Bitsanis1988_0}, which compute a local viscosity via the local density, 
our approach incorporates the true nonlocal nature of the viscosity by starting from microscopic equations of motion, and allows analysis in nonlinear situations, i.e., including the back-reaction of flow on the density distribution.

In a previous work, Ref.~\cite{AerovKrugerJCP2014}, we noted that using the Smoluchowski equation with a naive driving (shear) profile can lead to inhomogeneous stresses, and therefore unphysical solutions. In Ref.~\cite{AerovKrugerJCP2014}, we suggested adjustment of the driving profile such that homogeneous stresses are obtained (``stress ensemble''). Here, we start from the Smoluchowski equation including hydrodynamic interactions, where a driving profile as such does not exist, being replaced by the prescribed driving velocities of a set of non-Brownian particles (e.g. the plates of a rheometer). In this setup, no unphysical solutions of the Smoluchowski equation arise, as all forces are balanced properly from the beginning. This important insight is accompanied by the explicit demonstration that results found from the stress ensemble agree exactly with those found from inclusion of hydrodynamic interactions to leading order in the hydrodynamic radius of the particles. 

The paper is organized as follows. In Sec.~\ref{sec:Setup}, we introduce the studied system and give the Smoluchowski equation for the considered setup (Eq.~\eqref{Smoluchowski}), as well as the resulting general friction forces (Eq.~\eqref{eq:F}). In Sec.~\ref{sec:I}, we make these equations tractable with density functional theory by integrating out $N-2$ particle positions, and obtain the main equations of the paper, Eqs.~\eqref{ren} and \eqref{af}. Specific results are obtained in Sec.~\ref{sec:R}, where we study the case of two walls sheared with respect to each other. We conclude in Sec.~\ref{sec:C}.

\section{Setup and equations of motion}\label{sec:Setup}
\subsection{Setup}
Consider $N$ Brownian particles (BPs) and $n$ non-Brownian particles (nBPs) immersed in a solvent (Fig.~\ref{fig:1}, left-hand side). 
The nBPs play the role of the (moving) confinement, their positions and velocities are controlled from outside. The main goal of the paper is to find the friction forces acting on the nBPs on their predefined trajectories. This will yield the rheological properties of the confined suspension (viscosity), depending among others on size, shape, position, and velocity of the nBPs.  

The setup encompasses many realistic situations, e.g. cases termed {\it microrheology} \cite{rauscher2_0, Brady_Squires,WilsonEPL2011_Microrh_0}, when a small nBP is driven through the suspension of BPs having a comparable size; It also comprises  the case of two walls moving at a distance comparable to the size of the BPs, as discussed in detail in Sec.~\ref{sec:R} below (see the right hand side of Fig.~\ref{fig:1}).  

We note that, strictly speaking, the described setup does not include other cases of microrheology, were the diffusion or sedimentation of tracer particles \cite{batchelor72,batchelor83,Beenakker84a} is studied. In those cases, the (external) driving is invoked by forces, rather than by the motion of nBPs considered here. 

\begin{figure*}
\includegraphics[width=0.9\textwidth]{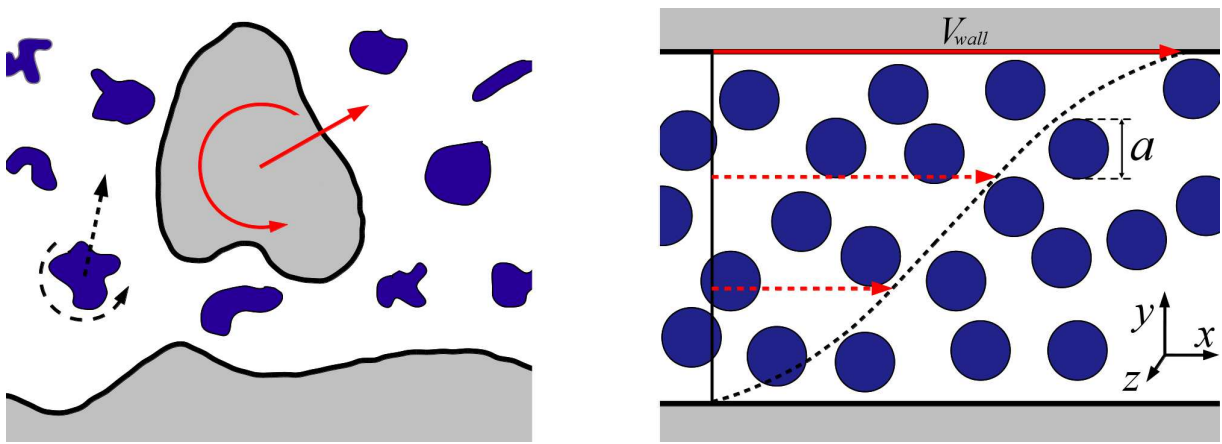}
\caption{\label{fig:1}(Color online) Left: Suspension consisting of (gray, solid arrows) non-Brownian particles that are controlled from outside, and (blue, dashed arrows) Brownian particles. The goal of this article is to compute the (friction) forces acting on the nBPs, which are a measure for the viscosity of the suspension. In section \ref{sec:Setup}, we give the general Smoluchowski equation for the BPs as well as the forces acting on the nBPs, valid for any shape of the involved particles (Eqs.~\eqref{Smoluchowski} and \eqref{eq:F}). Right: The specific example case studied in Sec.~\ref{sec:R}; a suspension of spherical Brownian particles sheared between two walls (where the nBPs take the role of the walls).} 
\end{figure*}

\subsection{Equations of motion -- Exact Smoluchowski equation}
We start by considering the setup on the lhs of Fig.~\ref{fig:1} in full generality. Thus, the vector $\vct{R}$ (in general $6(n+N)$ dimensional, due to $3$ translational and $3$ rotational degrees of freedom) denotes the particle positions and orientations, and $\vct{V}\equiv \partial_t \vct{R}$ are the corresponding velocities (including translation and rotation). Restricting to laminar flow,
hydrodynamic interactions (HI) are linear in the velocities, and instantaneous on the time scales considered \cite{dhont}. 
The hydrodynamic force $\vct{F}^h$ acting on the $(n+N)$ particles are found 
from the friction matrix $\mathbb{G(\vct{R})}$, \cite{dhont}, depending on all particle coordinates,
\begin{align} \label{eq:FGV}
\vct{F}^h= -\mathbb{G}\vct{V}.
\end{align}
Eq.~\eqref{eq:FGV} explicitly displays the linearity of laminar hydrodynamic flow.

One can now regard the subset $\vct{F}^h_N$, i.e., the hydrodynamic forces acting on the BPs, by projecting Eq.~\eqref{eq:FGV} on that $6N$ dimensional subspace. Also explicitly splitting the vector $\vct{V}$ into the two subsets, we obtain
\begin{multline} \label{hydrForBP}
\vct{F}^h_N(t)= -\mathbb{G}_{\bf Nn}[\vct{R}_{\mathbf{n}}(t),\vct{R}_{\mathbf{N}}(t)]\vct{V}_{\mathbf{n}}(t)-\\
\mathbb{G}_{\bf NN}[\vct{R}_{\mathbf{n}}(t),\vct{R}_{\mathbf{N}}(t)]\vct{V}_{\mathbf{N}}(t).
\end{multline}
Here, we have introduced subscripts that denote dimensionality, which will be used in the following. E.g. $\vct{V}_{\mathbf{N}}$ spans the subspace of BPs; 
$\mathbb{G}_{\mathbf{Nn}}$ is a matrix transforming from the nBPs' to the BPs' subspace. In Eq.~\eqref{hydrForBP}, we have also explicitly given the dependence of $\mathbb{G}$, being functions of all particle positions at time $t$.

On the Brownian time scale \cite{dhont}, momenta of BPs are relaxed, and the Smoluchowski equation follows from balancing forces acting on the BPs.
Additionally to the hydrodynamic force in Eq.~\eqref{hydrForBP}, there is the so-called Brownian force \cite{dhont} due to thermal fluctuations, 
\begin{align} \label{BrownianForces}
\vct{F}^{thermal}_N(t)=- k_BT\frac{\partial \ln P(\vct{R}_{\mathbf{N}},t) } { \partial \Rb_{\mathbf{N}}}, 
\end{align}
where $P(\vct{R}_{\mathbf{N}},t)$ is the time dependent probability distribution of the BPs. 
Each BP is also subject to potential forces exerted by all other particles,   
\begin{align} \label{PotentialForce}
\vct{F}^{potential}_N(t)=- \frac{\partial W(\vct{R}(t)) } { \partial \Rb_{\mathbf{N}}}, 
\end{align}
where $W(\vct{R}(t))$ is the interaction potential of all particles in the system. 
Balancing these forces leads to  
\begin{align} \label{ForceBalance}
0=\vct{F}^{potential}_N(t)+\vct{F}^{thermal}_N(t)+\vct{F}^h_N(t). 
\end{align}

By substituting Eqs.~\eqref{hydrForBP},~\eqref{BrownianForces},~\eqref{PotentialForce} into Eq.~\eqref{ForceBalance}, and multiplying the resulting equation 
by $\mathbb{M}\equiv[\mathbb{G}_{\bf NN}]^{-1}$, we get the velocity of the BPs 
\begin{align}\label{eq:v}
\vct{V}_{\mathbf{N}}= -\mathbb{M} \left( \frac{\partial} { \partial \Rb_{\mathbf{N}}}  \left[W+k_BT \ln P \right] +  \mathbb{G}_{\mathbf{Nn}}\vct{V}_{\mathbf{n}}\right).
	\end{align}
We note that $\mathbb{M}$, the mobility matrix for the Brownian subspace, can be identified as the mobility matrix for the case where the nBPs 
are static (at rest),
\begin{align}\label{eq:2}
	        \vct{V}_{\mathbf{N}}= -\mathbb{M} \vct{F}_{\mathbf{N}}^{h}, \hspace{1mm} \mbox{ if } \vct{V}_{\mathbf{n}}\equiv0.
	\end{align}
$\mathbb{M}$ is thus a well defined matrix, being $6N$ dimensional, however still depending on the positions of all particles \footnote{Note that $\mathbb{M}$ is different from $[\mathbb{G}^{-1}]_{\bf NN}$, the latter being obtained from projection on the subspace $\bf N$ {\it after} inversion. $[\mathbb{G}^{-1}]_{\bf NN}$ plays not role for our analysis.}.

The Smoluchowski equation is found from continuity, \cite{archer, dhont}, i.e., $\frac{\partial}{\partial t}P\!=\!- \frac{\partial}{\partial \Rb_{\mathbf{N}}} \!\! \cdot \!\! \vct{V}_{\mathbf{N}}P$, 
	
\begin{align}\label{Smoluchowski}
\frac{\partial}{\partial t}P\!=\!\frac{\partial}{\partial \Rb_{\mathbf{N}}}   \! \cdot \mathbb{M} 
\left(  \frac{\partial}{\partial \Rb_{\mathbf{N}}} \left[W\!\!+\!\!k_BT \ln P\right] +  \mathbb{G}_{\mathbf{Nn}}\vct{V}_{\mathbf{n}}\right)P .
\end{align}
Eq.~\eqref{Smoluchowski} yields $P(\vct{R}_{\mathbf{N}}, 
t)$, by itself a quantity of interest, measurable e.g. by confocal microscopy. With it, any (time dependent) observable is accessible in this
framework, e.g.~mean squared displacements. We focus on the generalized friction forces $\vct{F}_{\mathbf{n}}$ acting on the nBPs. 

We note that Eq.~\eqref{Smoluchowski} does not contain a mean solvent flow velocity, which is in contrast to commonly studied cases including driving flow, but neglecting HI, as e.g. Eq.~(1) in Ref.~\cite{AerovKrugerJCP2014}. We also note that Eq.~(1) in Ref.~\cite{AerovKrugerJCP2014}, for the case of shear, does not follow from Eq.~\eqref{Smoluchowski} by taking the leading order in HI.

The BPs' velocities are a function of their positions and the distribution $P$, see Eq.~\eqref{eq:v}. 
The mean of $\vct{F}_{\mathbf{n}}$ on the Brownian time scale, is hence 
\begin{align}
	\langle\vct{F}_{\mathbf{n}}^{}\rangle(t)\!\!=\!\!-\!\!\int \!\! d\vct{R}_{\mathbf{N}} P(t) \! \left[\mathbb{G}_{\mathbf{nn}} {\bf V}_{\mathbf{n}}
	\!\!+\!\mathbb{G}_{\mathbf{nN}}
	{\bf V}_{\mathbf{N}}\!\!+\!\!\frac{\partial} { \partial \Rb_{\mathbf{n}}} W\right]\!\!.\label{eq:F}
\end{align}
The first term on the rhs of Eq.~\eqref{eq:F} is the force induced by the motion of the nBPs. The second
term contains the force on the nBPs due to the motion of BPs. The last term represents the potential force.

The friction force in Eq.~\eqref{eq:F} is the force acting on the moving nBPs (or moving boundaries), which is a measurable  and relevant quantity. Finding this force is the main goal of this manuscript, as it is a measure of the viscosity of the confined suspension of BPs. Note that in this setup, we do not have immediate access to local quantities like stress (in contrast to the stress ensemble model discussed in Sec.~\ref{sec:ens}) or stresslets \cite{KimKarrila}. 

$\langle\vct{F}_{\mathbf{n}}^{}\rangle(t)$ depends generally on the trajectories of nBPs in the past. Eq.~\eqref{eq:F} gives the mean force, but higher moments,
e.g. force fluctuations~\cite{Brady_Squires}, are also accessible once $P$ is known.

Eqs.~\eqref{Smoluchowski} and \eqref{eq:F} are valid for arbitrarily shaped BPs and nBPs. (Analytical) Analysis is challenging in general, and exact solutions have mostly been restricted to small $N$ and $n$, see e.g.~Ref.~\cite{Brady_Squires,rauscher2_0} for the case of a nBP dragged through a suspension of BPs.
In the following section, we  proceed by making Eqs.~\eqref{Smoluchowski} 
and \eqref{eq:F} amenable to (approximate) treatments via classical DFT \cite{bob_advances}.

\section{Integrating out particles and density functional theory}\label{sec:I}
\subsection{Pairwise potential}
In this section, we restrict to spherical Brownian particles which interact via the pairwise potential $\phi(\rb_{ij})$, depending only on the respective center-center-distance $\rb_{ij}$. More specifically, denoting $\rb_{i}$ 
the coordinate of particle $i$, we split the potential into a term depending only on the nBPs ($U$), a term describing the pairwise interaction between a BP and the nBPs ($V$), and $\phi(\rb_{ij})$,  
\begin{align} \label{pairwise_potential}
W(\Rb)=U(\Rb_\mathbf{n})+\sum_{i=1}^N V(\rb_i, \Rb_\mathbf{n})+\sum_{j\neq i}\sum_{i=1}^N \phi(\rb_{ij}).
\end{align}
For the following integration procedure, it is irrelevant whether $U(\Rb_\mathbf{n})$ or $V(\rb_i, \Rb_\mathbf{n})$ are pairwise for nBPs as we do not integrate over their coordinates. In order to be able to integrate Eqs.~\eqref{Smoluchowski} and \eqref{eq:F} over $N-1$ or $N-2$ (Brownian) particle positions (see e.g. \cite{archer}), we also have to restrict to pairwise hydrodynamic interactions, which simplifies the matrices $\mathbb{G}$ and $\mathbb{M}$, as specified in the next subsection.
\subsection{Expansion of the hydrodynamic tensors}\label{Expansion}
Here we expand the friction and mobility tensors into components depending on one, two, three, \dots \! BPs, respectively. When arriving at Eqs.~\eqref{ren} and \eqref{af} below, we keep only those matrices depending on one or two BPs. This yields the leading order (pairwise) HI. For the special case of spheres, such series expansions can be assumed to converge if the hydrodynamic diameter $a_H$ is smaller than the interaction diameter $a$. 

	The mentioned expansion can be done in a well defined manner; Let us give as an exemplary case the mobility matrix $\mathbb{M}$ using Eq.~\eqref{eq:2}, while the remaining tensors are expanded in Appendix \ref{sec:AH}. Let $\Fb_j^h$ be the  hydrodynamic force 
for BP number $j$ (i.e., $\Fb_j^h$ is the $j$th  part of $\Fb_N^h$), and the position of this particle is $\rb_j$. Then we have for  its velocity,
\begin{multline} \label{M_definition}
-\left[\mathbb{M}(\Rb)\Fb_N^h\right]_{j}\equiv\\
-\mathbb{M}_{\mathbf{11}}^{(1)}(\rb_{j}, \Rb_n)\Fb_{j}^h-
\sum_{k \neq j}^N\mathbb{M}_{\mathbf{12}}^{(2)}(\rb_{j},\rb_k, \Rb_n)\left(\Fb_{j}^h,\Fb_k^h\right)^T+\dots,
\end{multline}
This defines the tensors $\mathbb{M}_\mathbf{11}^{(1)}(\rb, \Rb_n)$ and $\mathbb{M}_{\mathbf{12}}^{(2)}(\rb,\rb', \Rb_n)$ used below in Eqs.~\eqref{ren} and \eqref{af}. Note that $\left(\Fb_{j}^h,\Fb_k^h\right)$ is a line vector (i.e., a part of $\Fb_N^h$ in Eq.~\eqref{eq:2}), and Eq.~\eqref{M_definition} is still linear in forces. Recall that indices of $\mathbb{M}$ denote dimensionality (not particle index), so that e.g.~$\mathbb{M}_\mathbf{11}^{(1)}$ refers to the subspace of one Brownian particle.   

\subsection{Integrating out particle positions}\label{Integration}
In this subsection, we give the main steps necessary to perform the integration over $N-1$ or $N-2$  particle positions, see also Ref.~\cite{archer}.  Let $\mathbb{O}(\rb_1)$ denote a hydrodynamic tensor depending on BP position $\rb_1$ (e.g. the first term of the matrix expansions of Subsection~\ref{Expansion} and Appendix \ref{sec:AH}).  
We then have in Eqs.~\eqref{Smoluchowski} and~\eqref{eq:F} terms reading like $\mathbb{O}(\rb_1) P\left(\Rb_N,t\right)$. For these,  the well known exact integration over $N-1$ particles can be performed,  
\begin{align} \label{rho1_integration}
N \int \mathbb{O}(\rb_1) P\left(\Rb_N,t\right) d\rb_2...d\rb_N =\rho(\rb_1,t)\mathbb{O}(\rb_1),
\end{align}
where the one body density appears \cite{archer, HansenMcDonald}
\begin{align} \label{rho1_definition}
\rho(\rb_1,t)\equiv N \int P\left(\Rb_N,t\right) d\rb_2...d\rb_N .
\end{align}
Terms involving a hydrodynamic tensor $\mathbb{O}(\rb_1, \rb_2)$ depending on two positions (or containing the pairwise potential $\phi)$, can only be integrated over $N-2$ positions, 
\begin{multline} \label{rho2_integration}
N (N-1)\int \mathbb{O}(\rb_1, \rb_2) P\left(\Rb_N,t\right) d\rb_3...d\rb_N =\\
\rho^{(2)}(\rb_1,\rb_2, t)\mathbb{O}(\rb_1, \rb_2).
\end{multline}
Here, the two-particle density of BPs enters, 
\begin{align} \label{rho2_definition}
\rho^{(2)}(\rb_1,\rb_2, t)\equiv N (N-1) \int P\left(\Rb_N,t\right) d\rb_3...d\rb_N.
\end{align}
One subtlety arises in the integration procedure as two different types of interactions are present. Although interactions are pairwise, a BP can interact via $\phi$ with a second one, which in turn can interact via HI with a third one. This introduces also $\rho^{(3)}$, which is however in a suitable form for a well known identity, connecting it to $\rho^{(2)}$ (strictly valid in equilibrium) \cite{HansenMcDonald,RexLoewenPRL2008}, see Appendix \ref{AI}.
\subsection{Resulting equations}
Using the expressions presented in Subsections~\ref{Expansion} and~\ref{Integration} and the corresponding appendices, we finally obtain by integrating Eq.~\eqref{Smoluchowski} over $N-1$ particle positions
\begin{align} \label{rho1_evolution}
\frac{\partial \rho (\rb_1,t) }{\partial t}=\frac{\partial}{\partial \rb_1} \cdot \int d \rb_2 \Jb (\rb_1, \rb_2,t),
\end{align}
where we introduced the two-particle current $\Jb$,
\begin{widetext}
\begin{align} \label{ren}
	&\Jb(\rb_1,\rb_2,t)\equiv\delta(\rb_2) {\bf j}^{(1)}(\rb_1,t)+{\bf j}^{(2)}(\rb_1,\rb_2,t)\notag,\\   
 &\equiv\delta(\rb_2){\rho(\rb_1,t)}
	\mathbb{M}^{(1)}_{\mathbf{11}}(\rb_1)\left[\tilde\Fb_{\mathbf{1}}(\rb_1,t)+\mathbb{G}_{\mathbf{1n}}^{(1)}(\rb_1)\Vb_{ \mathbf{n} }\right]+\rho^{(2)} (\rb_1,\rb_2,t)\left[\mathbb{M}_{\mathbf{12}}^{(2)}(\rb_1,\rb_2)\tilde\Fb_{\mathbf{2}}(\rb_1, \rb_2, t) +\left(\mathbb{M}\mathbb{G}\right)_{\mathbf{1n}}^{(2)}(\rb_1,\rb_2)\Vb_{ \mathbf{n} }\right]\notag.\\
\end{align}
Here, we introduced  ${\bf j}^{(1)}$ and ${\bf j}^{(2)}$, which allows a compact representation of the force in Eq.~\eqref{af} below. $\tilde\Fb_{\mathbf{1}}(\rb ,t)$ is an auxiliary function, being an effective one body force acting on a BP at position $\rb$ \cite{archer}, 
\begin{align}\label{eq:aF}
\tilde\Fb_{\mathbf{1}}(\rb ,t) \equiv  k_BT  \rho^{-1}(\rb,t) \frac{\partial}{\partial \rb} \rho (\rb,t)+
\frac{\partial}{\partial \rb} V(\rb ,\Rb_{ \mathbf{n} })
+\rho^{-1}(\rb,t)\int d \rb' \rho^{(2)}(\rb,\rb',t) \frac{\partial}{\partial \rb} \phi(\rb-\rb'),
\end{align}
and $\tilde\Fb_\mathbf{2}(\rb_1, \rb_2, t) \equiv (\tilde\Fb_{\mathbf{1}}(\rb_1,t), \tilde\Fb_{\mathbf{1}}(\rb_2,t))^T$ is a six dimensional vector. We note that, importantly, by removing all nBPs from the system, Eqs.~\eqref{rho1_evolution} and \eqref{ren} can be identified with Eq.~(2) in Ref.~\cite{RexLoewenPRL2008}, where nonequilibrium systems, however without externally applied flow, are studied with dynamical DFT.

The force acting on the nBPs on the level of $\rho^{(2)}$ is obtained similarly, by integrating Eq.~\eqref{eq:F}, 
\begin{align}\label{af}
	\langle{\bf F}_{\mathbf{n}}\rangle&=\notag{\bf F}_{\mathbf{n}}^{0} 
       +\int  d \rb_1 d \rb_2 \Biggl[ \delta(\rb_2)\rho(\rb_1,t)\left[\nabla V(\rb_1)
       -\mathbb{G}^{(1)}_{\mathbf{nn}}(\rb_1)\Vb_\mathbf{n}\right]+
       \mathbb{G}^{(1)}_{\mathbf{n1}}(\rb_1)\Jb(\rb_1,\rb_2,t)
       +\\&+\rho^{(2)}(\rb_1,\rb_2,t)\left(\mathbb{G}^{(2)}_{\mathbf{n1}}(\rb_1,\rb_2)\left[\frac{{\bf j}^{(1)}(\rb_1,t)}{\rho (\rb_1,t)}+{\bf j}^{(2)}(\rb_1,\rb_2,t)\right]
	-\mathbb{G}^{(2)}_{\mathbf{nn}}(\rb_1,\rb_2)\Vb_\mathbf{n}\right)\Biggr].
\end{align}
\end{widetext}
${\bf F}_{\mathbf{n}}^{0} \equiv -\frac{\partial} { \partial \Rb_{\mathbf{n}}} U(\Rb_\mathbf{n}) - \mathbb{G}^{(0)}_{\mathbf{nn}}\Vb_{\mathbf{n}}$ 
denotes the force in absence of BPs, were $\mathbb{G}^{(0)}_{\mathbf{nn}}$ is the matrix describing the situation in absence of BPs, see Eq.~\eqref{Gnn_definition}. (This force does not fluctuate in our framework, and no averaging is needed.)  

The integrated Smoluchowski equation for shear without hydrodynamic interactions, see e.g. Eqs.~(3) and (4) in Ref.~\cite{AerovKrugerJCP2014}, follows from Eq.~\eqref{ren} by replacing $\mathbb{M}^{(1)}_{\mathbf{11}}k_BT=D_0$ and 
$-\mathbb{M}^{(1)}_{\mathbf{11}}\mathbb{G}^{(1)}_{\mathbf{1n}}\vct{V}_{\mathbf{n}}=\vct{V}$, and neglecting all tensors with superscript 2. Then, $\vct{V}$ is the solvent velocity induced by the moving nBPs and $D_0$ is the bare BPs' diffusivity. We note that Eqs.~(3) and (4) in Ref.~\cite{AerovKrugerJCP2014} (as mentioned above already) cannot easily be derived from the more precise Eq.~\eqref{ren}, e.g. by taking the limit of weak HI. (It is because the shear term in Eq.~(4) in Ref.~\cite{AerovKrugerJCP2014} is not of the same order in HI as the remaining terms in that Eq.~(4).) 

Using Eqs.~(3) and (4) in Ref.~\cite{AerovKrugerJCP2014}, we noted an inconsistency for cases of confinement, i.e., an inhomogeneous local shear stress, which is unphysical, and which we suggested to remove by use of the stress ensemble. In the latter the flow velocity in Eq.~(4) in Ref.~\cite{AerovKrugerJCP2014} is adjusted to obtain stress homogeneity. It is important to note that Eq.~\eqref{ren} in contrast indeed yields physical results throughout, as from the very beginning, all forces are balanced properly. This is one main insight gained from the present work through the inclusion of HI.

{\it Summarizing this section}, we obtained, by integration, an equation for $\rho$ and $\rho^{(2)}$, Eq.~\eqref{ren}, that is valid for spherical BPs that interact with a pairwise potential and pairwise hydrodynamic interactions, and arbitrary nBPs. We also computed the friction force acting on the nBPs, Eq.~\eqref{eq:F}, on the same level of accuracy. Since, apart from the mentioned limitations (e.g. pairwise interactions) these equations are exact, they should naturally include known specific cases that have been derived using the same limitations \cite{KimKarrila}. These include e.g. the microrheology cases studied in Refs.~\cite{Brady_Squires,rauscher2_0}, where the tracer particle constitutes the non-Brownian particle. In general, Eqs.~\eqref{ren} and \eqref{af} yield exact results for small densities of BPs. Although designed for confined systems, Eqs.~\eqref{ren} and \eqref{af} contain also bulk properties \cite{batchelor77,brady_morris}, such as e.g. the Einstein coefficient for the viscosity of dilute suspensions (see Appendix~\ref{A:E} for more details), although not directly, e.g. only when taking the moving boundaries far way from each other (this is shown explicitly in Figs.~\ref{fig:2} and \ref{fig:3}). For bulk systems, many body hydrodynamic interactions have been taken into account in Ref.~\cite{Beenakker84}, which thus goes beyond Eqs.~\eqref{ren} and \eqref{af}.

\section{Results for two parallel walls in relative motion}\label{sec:R}
\subsection{Setup and closure}
We finally study the explicit case of two parallel walls in relative motion (see the right hand side of Fig.~\ref{fig:1} or the inset of Fig.~\ref{fig:2}), a scenario accessible by experiments and simulations
\cite{Jabbarzadeh1997,  Petravic2_2006_JCP_viscosity_lin_resp_between_planes, PeylaEPL2011expSuspInSlit}. The lower wall is positioned  at $y=0$, the upper wall at $y=d$. 
The upper wall moves deterministically with time independent velocity $v$ in direction $x$, while the lower is at rest, defining a bare shear rate of $\dot\gamma_0=\frac{v}{d}$. We consider in the following the steady state, time independent situation which is assumed to be approached a sufficiently long time after the shear is started. Then the upper wall is subject to the time independent, generalized friction force $\vct{F}^{(u)}$ found from Eq.~\eqref{af}.  Its $y$ and $x$ components yield respectively the orthogonal pressure and the effective shear viscosity. We focus on the latter, and define the effective viscosity $\nu_{\rm eff}$,
\begin{align}\label{eq:vis}
	        \nu_{\rm eff}\equiv-\frac{F_x^{(u)}}{A\dot\gamma_0}.  
	\end{align}
	$A$ is the surface area of the wall.
	
	In this situation, the one body density is a function of $y$ only, $\rho(\vct{r})=\rho(y)$.  To solve Eqs.~\eqref{ren} and \eqref{af} and compute $F_x^{(u)}$, one must express $\rho^{(2)}$ (approximately) in terms of $\rho$. As shown in Ref.~\cite{AerovKrugerJCP2014},
	a simple superposition closure involving the distorted bulk pair distribution
	$g_{neq}(\rb)\equiv g(\rb)-g_{eq}(\rb)$ under shear suffices to capture the shear induced distortion of $\rho^{(2)}$ \cite{AerovKrugerJCP2014, Brader_Kruger_2011},
	\begin{align}\label{cl}
		\rho^{(2)}(\rb,\rb')&\approx \rho^{(2)}_{ad}(\rb,\rb') + \rho(\rb)\rho(\rb')g_{neq}(\rb-\rb').
	\end{align}
	$\rho^{(2)}_{ad}$, the so-called adiabatic term expressed via the density functional by Eq.~\eqref{sum_rule},  is the main ingredient of dynamical DFT \cite{archer} \footnote{Eq.~\eqref{sum_rule} can only be used if $\rho^{(2)}$ appears in the form of the left hand side of Eq.~\eqref{sum_rule}. This is the case when evaluating Eq.~\eqref{ren} to leading order in HI, see e.g. the last term in Eq.~\eqref{eq:aF}. In general, when regarding higher orders in HI, other closures are necessary also for the ``adiabatic term'', e.g.  superposition approximations \cite{RexLoewenPRL2008}.}. 
	For our hard sphere system we use the Rosenfeld form of the excess part of the free energy $\mathcal{F}_{\rm ex}$. This term is essential, as it correctly captures the equilibrium structure of
	the fluid between the walls. However, it does not describe effects of shear \cite{Brader_Kruger_2011}, making the second term in Eq.~\eqref{cl} important for the considered (sheared) system. 
In Ref.~\cite{AerovKrugerJCP2014}, its properties are analyzed in detail.}	
	
	Eq.~\eqref{cl} is by construction exact in homogeneous systems, and it uses knowledge about bulk rheology \cite{joe_review}, imprinted in $g_{neq}(\rb)$, to describe inhomogeneous
	systems. In Ref.~\cite{AerovKrugerJCP2014}, we demonstrated that Eq.~\eqref{cl} yields the exact contact density (corresponding to the normal force exerted on the wall by the particles) for shear flow at a single wall, as well as the
	correct scalings for shear rate demanded by symmetry \footnote{E.g., the shear stress is linear in shear rate for small rates, while the orthogonal pressure changes quadratically.}.
	This framework, needing the closure in Eq.~\eqref{cl}, will also benefit from recent developments in dynamical
	DFT (``power functional'')\cite{SchmidtBraderJCP_2013_power_func, Brader13}.
	
	\subsection{Results from Eqs.~\eqref{ren} and \eqref{af} for small hydrodynamic radii}\label{sec:saH}
	Eqs.~\eqref{ren} and \eqref{af}  (with Eq.~\eqref{cl}) can in principle be evaluated to any accuracy 
of (pairwise) hydrodynamic interactions, and it is instructive to introduce BPs with hydrodynamic radius $a_H/2$ and hard interaction radius $a/2$,
as then, for $a_H<a$, convergence of a 
series in $a_H/a$ may be assumed. The corresponding matrices $\mathbb{G}$ and $\mathbb{M}$ for spherical particles between parallel walls can for example be found in Ref.~\cite{SwanBrady2010}, 
and $g_{neq}$ (in Eq.~\eqref{cl}) is given in Ref.~\cite{brady_morris} to any order in $a_H/a$. 

In zeroth order in $a_H/a$, BPs are infinitely fast, and $P$ adjusts instantaneously to the equilibrium distribution. This is explicitly found from Eq.~\eqref{rho1_evolution}, 
which requires then $\tilde\Fb_{\mathbf{1}}=0$, as fulfilled by $\rho_{eq}$ \cite{HansenMcDonald}.
In this order, $\nu_{\rm eff}$ equals the bare solvent viscosity $\nu_0$. In general,
\begin{align}\label{nu}
\nu_{\rm eff}=\nu_0+\nu_1 a_H+\nu_2 a_H^2 +\dots.
\end{align}
$\nu_1$, $\nu_2 $ and so on, depend on the distance $d$ as well as on velocity $v$ and average density. We will in the following analyze the first nontrivial term, $\nu_1$, and its dependence on distance $d$. It is worth noting that restricting to linear order in $a_H$ directly implies that the results are linear in the velocity $v$. 

For $\nu_1$  we still have that the solution for the one body density $\rho(y)$ equals the equilibrium one, the pair density $\rho^{(2)}$ is however distorted by the shearing. It is given by Eq.~\eqref{cl}, with the equilibrium form for the one body density, and, in the considered order in HI, we insert the form of Eq.~(33) of Ref.~\cite{AerovKrugerJCP2014} for $g_{neq}(\rb)$. 

With $\rho$ and  $\rho^{(2)}$ obtained in this manner, we use Eq.~\eqref{af} to obtain the force on the upper plate. In leading order $a_H$, all matrices with superscript 2 can be omitted, and we have   
\begin{multline}  \label{Force_Plane}
-\frac{\langle F_x^{(u)}\rangle}{A}=\dot\gamma_0\nu_0+\\
\frac{\hat{\bf{x}}}{A}\cdot\int d \rb \, \mathbb{G}^{(1)}_{\mathbf{n1}}(\rb)\mathbb{M}^{(1)}_{\mathbf{11}}(\rb)\int d \rb' \rho^{(2)}(\rb,\rb')\frac{\partial}{\partial \rb} \phi(\rb-\rb')=\\
\dot\gamma_0\nu_0+\int_0^d d y \, \frac{d-y}{d}\int d \rb' \rho^{(2)}(\rb,\rb')\frac{\partial}{\partial x} \phi(\rb-\rb')+\mathcal{O}(a_H^2).
\end{multline}
Note that here the shear distortion of $\rho^{(2)}(\rb,\rb')$ is evaluated to first order in $a_H$ (as mentioned), and that the equilibrium term for $\rho^{(2)}(\rb,\rb')$ (i.e., $\rho^{(2)}_{ad}(\rb,\rb')$ in Eq.~\eqref{cl}) does not contribute in the integral due to symmetries.
We also note that the force $F_x^{(u)}$ has no contribution from the one body density $\rho(\vct{r})$, as the result of cancellations in Eq.~\eqref{af}. To leading order in $a_H$, 
the effective viscosity is thus due to particle interactions, and there is no contribution from isolated particles, just as is the case for bulk systems, see e.g. \cite{brady_morris}. 

\begin{figure}
	\includegraphics[width=1\linewidth]{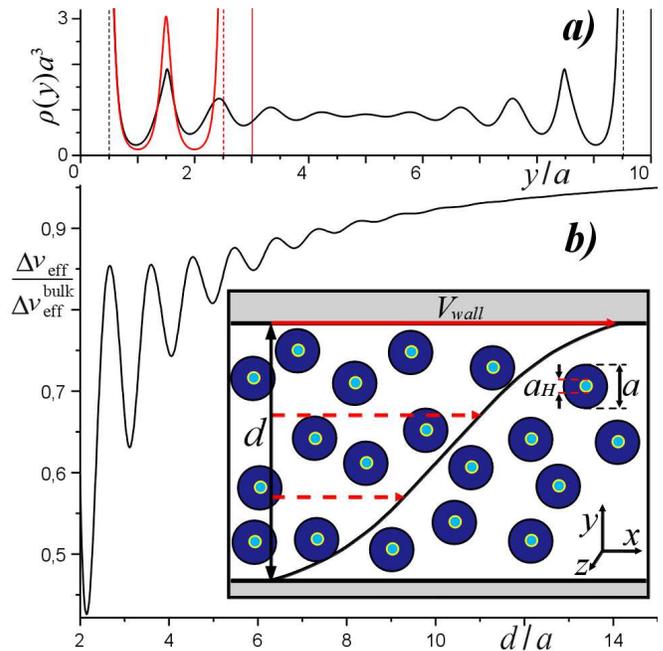}
	\caption{\label{fig:2}(Color online) b) Effective viscosity coefficient $\Delta \nu_{\rm eff}=\nu_{\rm eff}-\nu_0$ of a suspension sheared between walls, as a function of the distance $d$, 
	normalized to the bulk value. Packing fraction $\Phi=0.45$, and we consider particles with small hydrodynamic radius $a_H/2$ 
	and hard interaction radius $a/2$, $a_H\ll a$. 
	a) The corresponding equilibrium densities for two exemplary 
	cases, $d=10a$ and $d=3a$, \cite{DietrichPRE1997}.  Vertical dashed lines show the closest approach for particle centers.}
\end{figure}

Fig.~\ref{fig:2} b) shows the resulting viscosity for hard spheres confined by hard walls, for average packing fraction $\Phi=0.45$ 
(defined with respect to the interaction radius $a/2$), 
\footnote{In contrast to equilibrium cases 
\cite{bob_advances, DietrichPRE1997}, which are naturally discussed grand canonically, we prefer to keep the particle packing fixed, 
thus avoiding the definition of a chemical potential
in non-equilibrium.}, and small $a_H$, i.e. $a_H\ll a$. Specifically, the curve gives the coefficient $\nu_1$ in Eq.~\eqref{nu}, normalized by its bulk value
\footnote{For $d\to\infty$, $\nu_1= \frac{12\nu_0}{5a} \frac{\Phi^2 - \frac {\Phi^3}{2}}{(1-\Phi)^3}$. This expression, exact for small $\Phi$ (see also \cite{brady_morris}) is found from using Eqs.~(33) and (34) of Ref.~\cite{AerovKrugerJCP2014} for the distorted bulk pair correlation (again, exact for small $\Phi$ and $a_H$), and computing the bulk stress via Eq.~\ref{eq:sigmab}.}.
The curve approaching unity for large $d$ demonstrates that the present theory correctly finds the bulk limit.  
$\nu_1$ in tendency reduces to smaller values for decreasing $d$. While the curve is smooth for $d \agt 8a$, it develops oscillations for smaller $d$ 
due to layering effects, as seen in panel a). 

As mentioned before, $\nu_1$ is by construction linear in velocity $v$, and we will not study nonlinear effects in this subsection.

\subsection{Results in the stress ensemble model}\label{sec:ens}
\subsubsection{Stress ensemble model and its connection to Eqs.~\eqref{ren} and \eqref{af}}
While Sec.~\ref{sec:saH} and Fig.~\ref{fig:2} represent the case $a_H\ll a$, strictly following from expansion of Eqs.~\eqref{ren} and \eqref{af}, the observed qualitative scenario is possibly more general. In order to demonstrate this and to make connection to previous work, in this subsection, we use the model suggested in Ref.~\cite{AerovKrugerJCP2014} (where the case of a single wall was studied). In that model, we use the Smoluchowski equation with shear but without HI (see Eq.~(1) of Ref.~\cite{AerovKrugerJCP2014}), which means in the present framework to set (among others) $\vct{F}_{\mathbf{N}}^h D_0/k_BT= V(y)\hat{\bf x} -\vct{V}_{\mathbf{N}}$, with a solvent velocity $V(y)\hat{\bf x}$. As mentioned earlier, this can yield (unphysical) inhomogeneous shear stresses, i.e., the $xy$ component $\sigma_{xy}$ of the stress tensor ${\bf {\boldsymbol \sigma}}$ may depend on $y$. In absence of HI, the local stress tensor ${\bf {\boldsymbol \sigma}}$ is an exact functional of $\rho^{(2)}$ and $\rho$,
\begin{multline}
                \nabla \cdot {\bf {\boldsymbol \sigma}}(\vct{r})=-k_BT \nabla \rho(\rb)\\
                 - \int d^3\rb' [\frac{\partial}{\partial \rb} \phi(|\rb-\rb'|)] \rho^{(2)}(\vct{r},\vct{r}').\label{divsigma}
         \end{multline}
As in Ref.~\cite{AerovKrugerJCP2014}, we then adjust $V(y)$ until stress homogeneity, required by stationarity, is achieved. Specifically, we balance the total stress, made of particle contributions in Eq.~\eqref{divsigma}, and the stress from the solvent, $\nu_0 \frac {\partial^2 V(y) }{\partial y^2}$,
\begin{align}\label{sb}
\nu_0 \frac {\partial^2 V(y) }{\partial y^2}+\frac {\partial \sigma_{xy}}{\partial y}=0.
\end{align}
The particle stress in Eq.~\eqref{divsigma} follows unambiguously from the closure \eqref{cl}, and is hence self-consistently found \cite{AerovKrugerJCP2014}. See Appendix \ref{AS} for further details on the expression for the stress and its limit for bulk systems. 

It is interesting to note that the model proposed in Ref.~\cite{AerovKrugerJCP2014}, comprising Eqs.~\eqref{sb} and \eqref{divsigma},  exactly agrees with Eqs.~\eqref{ren} and Eq.~\eqref{af} for small $a_H$, as we aim to demonstrate; Integrating Eq.~\eqref{sb} twice, i.e., $\int_0^ddy\int_0^y dy'$, we obtain (using $\sigma_{xy}(y=0)=0$ \cite{AerovKrugerJCP2014} for hard particles),
\begin{align}\label{pr}
-\frac{\langle F_x^{(u)}\rangle}{A}= \nu_0\frac{\partial V}{\partial y}\bigr|_{y=0}=\dot\gamma_0\nu_0+\frac{1}{d}\int_{0}^{d}\sigma_{xy}(y)dy+\mathcal{O}(a_H^2).
\end{align}
Using Eq.~\eqref{average_stress}, this expression is identified with Eq.~\eqref{Force_Plane}, and hence the stress ensemble agrees with Eqs.~\eqref{ren} and \eqref{af} to leading order in $a_H$. The first equality in Eq.~\eqref{pr} follows because, knowing that the total stress is constant in space (from Eq.~\eqref{sb}), we only have to know it at one position, and choose the lower wall, where it is solely composed of (or carried by) the solvent. In order to arrive at Eq.~\eqref{pr}, we also identified $\dot\gamma=(V(d)-V(0))/d$ \footnote{Strictly, $\sigma_{xy}$ in Eq.~\eqref{Force_Plane} is computed for a simple shear profile, while $\sigma_{xy}$ in Eq.~\eqref{sb} is computed for a distorted flow profile. In leading (zeroth) order in $a_H$, the two are however identical.}.   

\subsubsection{Results linear in velocity $v$}
Having identified that the stress ensemble framework agrees with Eqs.~\eqref{ren} and \eqref{af} to leading order in $a_H$, we now use it for the case $a_H=a$, hoping that, due to its simplicity, 
it may capture generic features independent of system details (like HI).

\begin{figure}
	\includegraphics[width=1\linewidth]{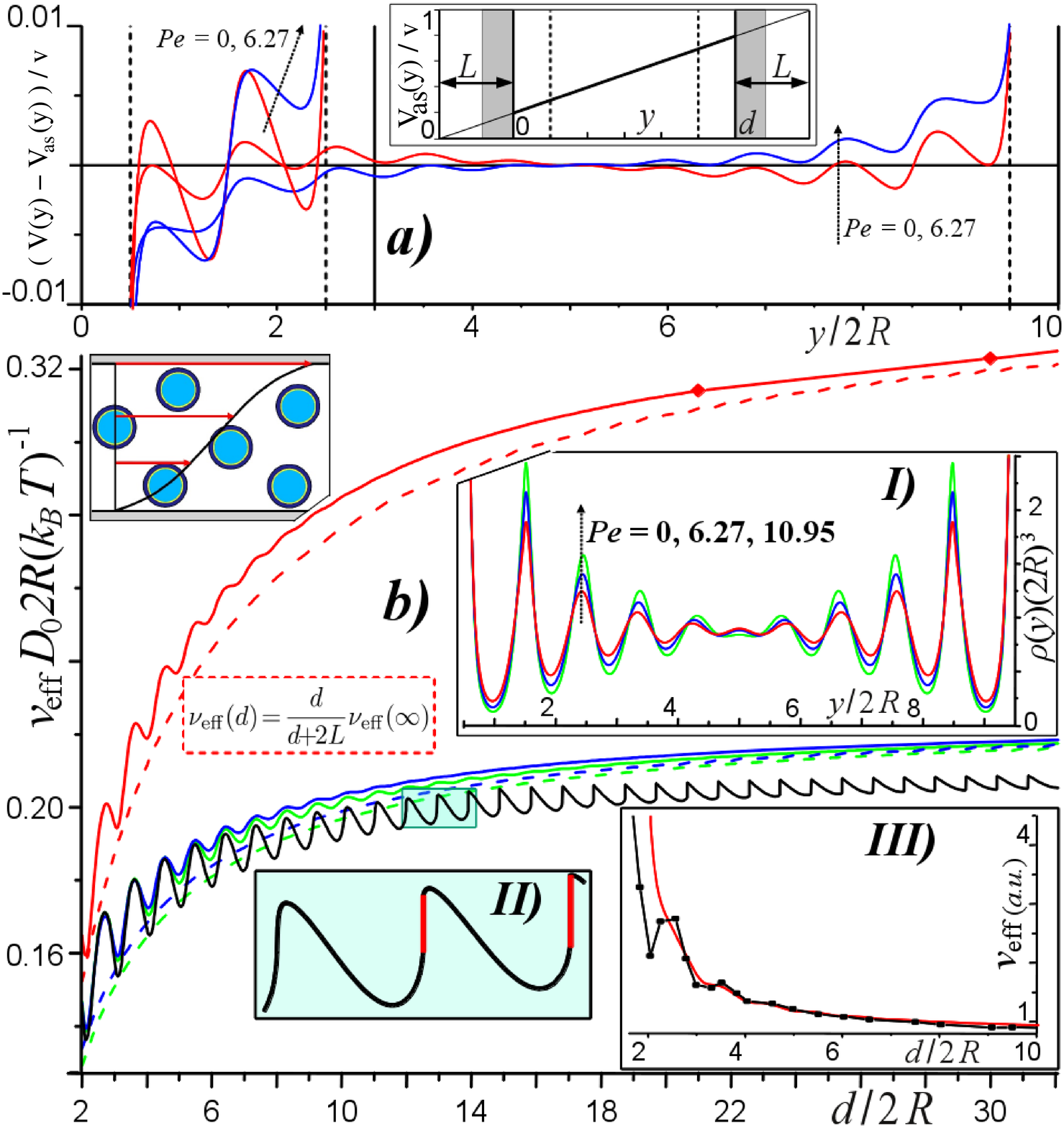}
	\caption{\label{fig:3}(Color online) b) Effective viscosity of a hard sphere fluid ($\Phi=0.45$, $a_H=a$) sheared between two walls as a function of distance $d$ for $Pe=0,6.27,10.95,20$ 
	(full curves, top to bottom). 
	Dashed curves are estimates based on slip effect, Eq.~\eqref{eq:est}. 
	I) Examples for density profiles under shear. II)  Enlarged segment of the $Pe=20$ curve, where 
discontinuities as a function of $d$ develop. III) Computer simulation results of Ref.~\cite{Jabbarzadeh1997}
fitted by the curve for $Pe=0$ assuming a different (negative) slip length, see main text. a) Corresponding velocity profiles for $d=3a$ and $d=10a$. The curves give the deviation of the flow velocity $V(y)$ from the flow profile approached for large $d$, denoted $V_{as}(y)$, normalized by the velocity of the upper wall $v$. Inset of a) shows the construction leading to the estimate of Eq.~\eqref{eq:est}.
}
\end{figure}

Fig.~\ref{fig:3} b) shows the resulting effective viscosity (Eq.~\eqref{eq:vis}) for different Peclet numbers $Pe\equiv \dot\gamma_0 a^2/D_0$, and $a_H=a$.
The upper curve shows the case of small $Pe$ (linear response), indeed possessing very similar features to 
Fig.~\ref{fig:2}: The effective viscosity approaches a distance independent bulk value for $d\to\infty$ which agrees exactly with the corresponding bulk result. Here, the bulk result is found again from using  $g_{neq}$ from Eq.~(33) of Ref.~\cite{AerovKrugerJCP2014} in formula \eqref{eq:sigmab}, \cite{brady_morris}. Specifically, this yields for $Pe \rightarrow 0$,
\begin{multline}
\nu_{\rm eff}(\infty)=\frac{k_BT}{3 \pi D_0 a} \left(1+\frac{12}{5} \frac{\Phi^2 - \frac {\Phi^3}{2}}{(1-\Phi)^3} \right).\\
\end{multline}
In the units used in Fig.~\ref{fig:3}, this corresponds to $\nu_{\rm eff}(\infty)=0.346 k_BT (D_0 a)^{-1}$ at $\Phi=0.45$.
Note again that this expression for the bulk viscosity, which indirectly also enters our results for confinement, is an exact solution of the Smoluchowski equation for small $\Phi$ and neglecting hydrodynamic interactions, and we used the Carnahan-Starling expression \cite{HansenMcDonald} 
to estimate the result for larger packing fractions.  

The effective viscosity in Fig.~\ref{fig:3} consistently decreases for small $d$. Again, for $d\alt 8a$, oscillations start to be visible. 

In Ref.~\cite{AerovKrugerJCP2014}, we computed the slip length $L$ of the suspension at a single wall under shear, e.g. $L=1.27a$ at $Pe\rightarrow0$. 
A simple geometric consideration (inset of Fig.~\ref{fig:3} a))
yields the following estimate for $\nu_{\rm eff}$ for two parallel walls at distance $d$, 
\begin{align}\label{eq:est}
	\nu_{\rm eff}(d)= \frac{d}{d+2L}\nu_{\rm eff}(\infty).
\end{align}
The outcome of Eq.~\eqref{eq:est} is shown in Fig.~\ref{fig:3} b) by dashed curves.
Despite the mentioned oscillations in the solid curve, which are not reflected by Eq.~\eqref{eq:est},
Eq.~\eqref{eq:est} gives an astonishingly good result even for small $d/a$. In particular, from Eq.~\eqref{eq:est}, one can estimate that the bulk value of the viscosity is approached with a power law of $1/d$ for large $d$.
	
In real systems, the effective viscosity depends on details, e.g. boundary conditions at the walls or particle dynamics. The simple picture following from
 our analysis identifies two main mechanisms, the slip-effect (Eq.~\eqref{eq:est}), determining the general behavior of effective viscosity as function of distance $d$ between the walls,
  and overlying oscillations with minima if $d$ is a multiple of the particle diameter.

\subsubsection{Results nonlinear in velocity $v$}
Upon increasing the driving velocity \footnote{For large Peclet numbers, we use Eq.~(34) of Ref.~\cite{AerovKrugerJCP2014} for the distorted bulk pair correlation, entering Eq.~\eqref{cl}.}, the asymptote for large $d$ decreases (lower curves in Fig.~\ref{fig:3} b)), which is due to the well known phenomenon of shear thinning in bulk systems at intermediate 
values of $Pe$. This thinning behavior is accompanied by more pronounced layering of the density for larger rates in inset I.
Apart from this, the overall qualitative features are very similar to the discussed cases, in particular, Eq.~\eqref{eq:est} gives a very good estimate of the overall trend 
for larger rates as well. Regarding $Pe=10.95$, we see that the oscillations in the viscosity extend to larger values of $d$. 
This is clearly a non-linear effect, as the higher rate causes changes in the density (see inset I of Fig.~\ref{fig:3} b)), which for increasing rates develops more pronounced oscillations,
extending to larger $d$.  

Bulk suspensions show layering at certain densities and shear rates, as found in simulations \cite{RastogiWagnerJCP1996LayeringSim_0, Foss00}, and confinement is then nontrivial. 
Our model \cite{Brader_Kruger_2011,AerovKrugerJCP2014}, i.e., usage of DFT with the closure Eq.~\eqref{cl}, predicts a layering instability at large $Pe$, i.e. oscillations of $\rho(y)$ 
for arbitrarily large $d$. The lowest curve of Fig.~\ref{fig:3} b) ($Pe=20$) representing such a state,
shows that the effective viscosity is {\it unsteady} as a function of $d$, having discrete jumps at $d\agt 12a$, which can be explained by the underlying density profiles.
At the jumps of $\nu_{\rm eff}(d)$, the density $\rho(y)$ is discontinuous as well, as the number of layers is changed by one. 
 The relative height of the discontinuities decays as $1/d$, since for larger $d$ each individual layer contributes less to the total viscosity.
 
 \subsubsection{Comparison to simulations}
 In the apparent absence of other data (experimental or from simulations) for suspensions at the considered densities,  the inset 
 III of Fig.~\ref{fig:3} b) shows simulation data for a {\it molecular fluid} \cite{Jabbarzadeh1997} sheared by two rough walls.

 The red curve (the smooth curve) is our result, i.e., the full red curve (the full top curve) of the main graph; in order to account for the different boundary conditions,
 we multiplied it by $(d+2 \times1.27 a)/(d-2\times0.84a)$, thus estimating the slip length in Ref.~\cite{Jabbarzadeh1997} as $L\simeq-0.84a$. 
 The curve reproduces well the overall 
 features of the simulation data.
 
The curve from Ref.~\cite{Jabbarzadeh1997} has oscillations (amplitude of which is of course detail-dependent) with minima roughly at $d$ equal to multiples of the particle diameter,
 as is the case for the prediction from our model. We may thus conclude, that although starting from a system of overdamped Brownian particles, the presented model captures the generic features also seen for molecular fluids. 
 
 

\section{Conclusions}\label{sec:C}	
	
{\it Physical:}  
Exemplified by monodisperse hard spheres, the viscosity of fluids  in confinement displays a variety of features. 
	For two walls in relative motion, the viscosity is astonishingly well described by a continuum estimate involving the slip length, Eq.~\eqref{eq:est},
	down to distances of a few particle diameters. According to the estimate, at large distances, the viscosity approaches the bulk value with a correction vanishing as a power law with $1/d$.
	At distances of a few particle diameters, the viscosity additionally displays oscillations as a function of distance showing local minima when $d$ is a multiple of the particle diameter. 
	At larger wall velocities, nonlinear effects are present, therefore the oscillatory behavior is extended to larger $d$.

{\it Technical:} We presented a formalism for analyzing the viscosity in confinement, designed for combination with Dynamical Density Functional Theory, starting from the Smoluchowski equation. The previously found inconsistency of the Smoluchowski equation with driving for inhomogeneous situations \cite{AerovKrugerJCP2014}, is healed when taking into account hydrodynamic interactions. We demonstrated that the previously suggested ensemble model \cite{AerovKrugerJCP2014} agrees exactly with the full hydrodynamic description to leading order in the hydrodynamic radius $a_H$.  

\begin{acknowledgments}
We thank J.~M.~Brader, R.~Evans, M.~Fuchs and J. Wu for useful discussions. This research was supported by Deutsche Forschungsgemeinschaft (DFG) grant No. KR 3844/2-1. 
\end{acknowledgments}
\begin{appendix}
	\section{Hydrodynamic tensors}\label{sec:AH}
	Let us consider the 
velocity component of BP number $i$ due to the velocities of the nBPs, $\Vb_n$, which is expanded as,

\begin{multline}  \label{MG_definition}
\left[\mathbb{M}(\Rb)\mathbb{G}_{\mathbf{Nn}}(\Rb)\Vb_n\right]_i\equiv\bigl[\mathbb{M}_{\mathbf{11}}^{(1)}(\rb_i, \Rb_n)\mathbb{G}_{\mathbf{1n}}^{(1)}(\rb_i,\Rb_n)+\\
\sum_{k \neq i}^N(\mathbb{MG})_{\mathbf{1n}}^{(2)}(\rb_i,\rb_k,\Rb_n)+\dots\bigl]\Vb_n.
\end{multline}
Together with \eqref{M_definition}, this  defines the tensors $\mathbb{G}_{\mathbf{1n}}^{(1)}(\rb,\Rb_n)$ and $(\mathbb{MG})_{\mathbf{1n}}^{(2)}(\rb,\rb',\Rb_n)$.

The hydrodynamic force for the nBPs due to the motion of nBPs is expanded in the following way, 
\begin{multline} \label{Gnn_definition}
-\mathbb{G}_{\mathbf{nn}}(\Rb_n)\Vb_n\equiv-\bigl[\mathbb{G}_{\mathbf{nn}}^{(0)}(\Rb_n)+\sum_{l=1}^N\bigl(\mathbb{G}_{\mathbf{nn}}^{(1)}(\rb_l,\Rb_n)+ \\
\sum_{k \neq l}^N\mathbb{G}_{\mathbf{nn}}^{(2)}(\rb_l,\rb_k,\Rb_n)+\dots\bigl)\bigl]\Vb_n,
\end{multline}
which is the definition of the other three tensors $\mathbb{G}_{\mathbf{nn}}^{(0)}(\Rb_n)$, $\mathbb{G}_{\mathbf{nn}}^{(1)}(\rb_l,\Rb_n)$, and $\mathbb{G}_{\mathbf{nn}}^{(2)}(\rb_l,\rb_k,\Rb_n)$;
In this special case the expansion starts with a component depending on no BP.

Finally, we expand the hydrodynamic force for nBPs due to the motion of BPs ($\Vb_i$ is the velocity of BP number $i$),
\begin{multline} \label{GnN_definition}
-\mathbb{G}_{\mathbf{nN}}(\Rb)\Vb_N\equiv-\sum_{l=1}^N\mathbb{G}_{\mathbf{n1}}^{(1)}(\rb_l, \Rb_n)\Vb_l- \\
\sum_{k \neq l}^N\mathbb{G}_{\mathbf{n{1}}}^{(2)}(\rb_l,\rb_k, \Rb_n)\left(\Vb_l, \Vb_k\right)^T+\dots,
\end{multline}
which is the definition of the tensors $\mathbb{G}_{\mathbf{n1}}^{(1)}(\rb_l, \Rb_n)$ and $\mathbb{G}_{\mathbf{n{1}}}^{(2)}(\rb_l,\rb_k, \Rb_n)$.

The so defined hydrodynamic tensors can be found in the literature for specific cases of BPs (e.g. for spheres with stick boundary conditions)  as well as specific shapes or arrangements of the 
nBPs \cite{KimKarrila}. (See e.g. Ref.~\cite{SwanBrady2010} for the case of spheres between parallel walls.)
\section{Position integration and free energy functional}\label{AI}
Even if restricting to pairwise hydrodynamic interactions and pairwise potential interactions $\phi$, 
we have the following terms in Eqs.~\eqref{Smoluchowski}, and~\eqref{eq:F}, that connect three different Brownian particles,
\begin{multline} \label{PhiRho3}
N (N-1)(N-2)\int \mathbb{O}(\rb_1, \rb_2) \phi(\rb_1, \rb_3) P\left(\Rb_N,t\right) d\rb_4...d\rb_N \\
=\rho^{(3)}(\rb_1,\rb_2, \rb_3, t)\mathbb{O}(\rb_1, \rb_2)\phi(\rb_1, \rb_3).
\end{multline}
This term can only be integrated over $N-3$ particles, and  the three body density appears,
\begin{multline} \label{rho3_definition}
\rho^{(3)}(\rb_1,\rb_2, \rb_3, t)\\
\equiv N (N-1) (N-2) \int P\left(\Rb_N,t\right) d\rb_4...d\rb_N.
\end{multline}
In order to eliminate  $\rho^{(3)}$, 
we use the following equilibrium relation, which derives from the second member of the
Yvon-Born-Green hierarchy (see e.g. Ref.~\cite{HansenMcDonald}), as also used in Ref.~\cite{RexLoewenPRL2008},
\begin{multline} \label{second YBG}
\int \frac{\partial \phi(\rb_1,\rb_2)}{\partial \rb_1} \rho^{(3)}(\rb_1,\rb_2,\rb_3)d \rb_3=-\bigl(k_BT\frac{\partial}{\partial \rb_1}+\\
\frac{\partial V(\rb_1, \Rb_n)}{\partial \rb_1} + \frac{\partial \phi(\rb_1,\rb_2)}{\partial \rb_1}\bigl)\rho^{(2)}(\rb_1,\rb_2).
\end{multline}
Here, $V$, as defined in Eq.~\eqref{pairwise_potential}, is the potential interaction between BPs and nBPs. 
We also use the so-called sum rule, see e.g. Ref.~\cite{archer},
\begin{multline}\label{sum_rule}
\int\!d\rb_2\, \rho^{(2)}(\rb_1,\rb_2)\frac{\partial}{\partial \rb_1}\phi(\rb_1, \rb_2)=\rho(\rb_1)\frac{\partial}{\partial \rb_1}\frac{\delta \mathcal{F}_{\rm ex}}{\delta \rho(\rb_1)},
\end{multline}
where $\mathcal{F}_{\rm ex}$ is the excess part of free energy (see e.g. Ref.~\cite{HansenMcDonald}). Note that Eqs.~\eqref{second YBG} and \eqref{sum_rule} are exact in equilibrium.
\section{Einstein viscosity}\label{A:E}
According to Einstein, the bulk viscosity for spherical particles with stick boundary conditions for the solvent is given by 
\begin{align}\label{eq:El}
	\nu_{\rm eff}/\nu_0=1+\frac{5}{2}\Phi+\mathcal{O}(\Phi^2),
\end{align}
where $\Phi$ is the packing fraction of spheres. This result should be contained in Eq.~\eqref{eq:F} as we aim to sketch briefly considering the case of two walls in parallel motion (see Fig.~\ref{fig:1} or Sec.~\ref{sec:R}). At small density of BPs, interactions between BPs and correlations in their positions can be neglected, their density between the walls being a constant number $\rho_0$ (hence, in this case, Eq.~\eqref{ren} is unnecessary). Eq.~\eqref{af} for the force reduces  to 
\begin{multline}\label{af_Einst}
       -\langle{\bf F}_{\mathbf{n}}\rangle= A\nu_0\dot\gamma_0 \hat{\bf{x}} + \\
       \rho_0\int \! d \rb \left[
       -\mathbb{G}^{(1)}_{\mathbf{nn}}(\rb)+\mathbb{G}^{(1)}_{\mathbf{n1}}(\rb)\mathbb{M}^{(1)}_{\mathbf{11}}(\rb)\mathbb{G}^{(1)}_{\mathbf{1n}}(\rb)\right]\Vb_\mathbf{n},
\end{multline}	
where $\hat{\bf{x}}$ is the unit vector in direction $x$. This is the exact description for the force necessary to shear a single sphere (or a dilute suspension of spheres) between two walls (when using the proper matrices for such geometry). It has been investigated numerically in Refs. \cite{SwanBrady2010,Sangani2011_confined_suspension_streesslet}, and the corresponding viscosity increment due to the presence of the sphere (related via Eq.~\eqref{eq:vis} to the force in \eqref{af_Einst}) was found to approach \eqref{eq:El} in the limit when the distance between the walls is large compared to the size of the sphere. 

Of course there are more direct ways of finding \eqref{eq:El} (which do not need the presence of the walls from the beginning), but in the framework of Eqs.~\eqref{ren} and \eqref{eq:F}, being designed for studying confined suspensions, the presence of nBPs is essential. 
\section{Stress tensor}\label{AS}
In Ref.~\cite{AerovKrugerJCP2014} we introduced the  local interparticle stress tensor, which obeys the exact relation 
\begin{multline}
\nabla \cdot {\bf {\boldsymbol \sigma}}(\vct{r})=-k_BT \nabla \rho(\rb)\\
 - \int d^3\rb' [\frac{\partial}{\partial \rb} \phi(|\rb-\rb'|)] \rho^{(2)}(\vct{r},\vct{r}').
\end{multline}

The stress tensor $ {\boldsymbol \sigma}$ itself is given by \cite{Kreuzer}, 
\begin{multline}
{\bf {\boldsymbol \sigma}}(\vct{r})=-k_BT\rho(\vct{r}) {\bf I}+\frac{1}{2}\int_0^1 d\lambda\int d^3\rb_1 \times \\
\times \frac{\vct{r}_1\vct{r}_1}{r_1} \left[\frac{\partial}{\partial r_1} \phi(r_1)\right]\rho^{(2)}(\vct{r}+(1-\lambda)\vct{r}_1,\vct{r}-\lambda\vct{r}_1) \:.\label{eq:sigmai}
\end{multline}
For homogeneous systems, this expression reduces to the well known expression, \cite{irving_kirkwood},
\begin{align}
{\bf {\boldsymbol \sigma}}=-k_BT\rho {\bf I}+\frac{1}{2}\rho_0^2\int d^3 \rb \frac{\vct{r}\vct{r}}{r} \left[\frac{\partial}{\partial r} \phi(r)\right]g(\vct{r}) \:,\label{eq:sigmab}
\end{align} 
where $\rho_0$ is the homogeneous (bulk) density of particles.

For the coordinate system depicted  in Fig.~\ref{fig:2}, we have, using translational invariance along $x$ and $z$,
\begin{equation} 
\frac {\partial}{\partial y}\sigma_{xy}(y)=-\int d^3\rb' \frac{x'}{r'} \left[\frac{\partial}{\partial r'} \phi(r')\right] \rho^{(2)}(\vct{r},\vct{r}+\vct{r}') \, .\label{eq:divsigmaSTRESS}
\end{equation}

Using Eq.~\eqref{eq:divsigmaSTRESS} one can transform Eq.~\eqref{Force_Plane} to 
\begin{align}
\label{average_stress}
-\frac{\langle F_x^{(u)}\rangle}{A}=\dot\gamma_0\nu_0+\frac{1}{d}\int_{0}^{d}\sigma_{xy}(y)dy+\mathcal{O}(a_H^2).
\end{align}

\end{appendix}


%

\end{document}